\documentclass{article}
\usepackage{spconf,amsmath,graphicx}

\usepackage{enumitem}
\setlist{nosep, leftmargin=*}

\usepackage{amssymb, amsfonts}
\usepackage{euscript}
\usepackage{mathtools}
\usepackage{booktabs}
\usepackage{multirow}

\usepackage{subcaption}
\usepackage[dvipsnames]{xcolor}
\usepackage{hyperref}
\graphicspath{Figures/}

\title{Attention Swin U-Net: Cross-Contextual Attention Mechanism for Skin Lesion Segmentation}

 \name{Ehsan Khodapanah Aghdam$^{\star}$ \qquad Reza Azad$^{\dagger}$ \qquad Maral Zarvani$^{\ddagger}$ \qquad Dorit Merhof$\:^{\S}$}

 \address{$^{\star}$ Department of Electrical Engineering, Shahid Beheshti University, Tehran, Iran \\ $^{\dagger}$ Institute of Imaging and Computer Vision, RWTH Aachen University, Aachen, Germany \\ $^{\ddagger}$ Department of Computer Engineering, Alzahra University, Tehran, Iran
 \\ $^{\S}$ Faculty of Informatics and Data Science, University of Regensburg, Regensburg, Germany}

\begin{document}
%
\maketitle
\begin{abstract}
Melanoma is caused by the abnormal growth of melanocytes in human skin. Like other cancers, this life-threatening skin cancer can be treated with early diagnosis. To support a diagnosis by automatic skin lesion segmentation, several Fully Convolutional Network (FCN) approaches, specifically the U-Net architecture, have been proposed. The U-Net model with a symmetrical architecture has exhibited superior performance in the segmentation task. However, the locality restriction of the convolutional operation incorporated in the U-Net architecture limits its performance in capturing long-range dependency, which is crucial for the segmentation task in medical images. To address this limitation, recently a Transformer based U-Net architecture that replaces the CNN blocks with the Swin Transformer module has been proposed to capture both local and global representation. In this paper, we propose Att-SwinU-Net, an attention-based Swin U-Net extension, for medical image segmentation. In our design, we seek to enhance the feature re-usability of the network by carefully designing the skip connection path. We argue that the classical concatenation operation utilized in the skip connection path can be further improved by incorporating an attention mechanism. By performing a comprehensive ablation study on several skin lesion segmentation datasets, we demonstrate the effectiveness of our proposed attention mechanism.
\end{abstract}
\section{Introduction} \label{sec:intro}
Automatic segmentation is a part of medical application domains, e.g., skin cancer. Skin cancer is one of the most contagious and fatal forms of cancer. The human skin forms from the three types of tissues in each layer as follows: The epidermis (top layer), dermis (middle layer), and hypodermis (bottom layer). While under the rational ultraviolet radiation from sunshine and the presence of melanocytes in the epidermis, it produces melanin at a considerably unusual rate. The lethal form of skin cancer is malignant melanoma resulting from the remarkable growth of melanocyte levels in the epidermis with a fatality rate of 1.62\%. According to the American Cancer Society, 99,780 new melanoma cases are expected by the end of 2022, with a mortality rate of 7.66\% (7,650 cases) \cite{siegel2022cancer}. Melanoma cancer's five-year survival rate drops from 99\% to 25\% if diagnosed at advanced stages due to its aggressive nature \cite{siegel2022cancer}, to this end, early diagnosis plays a crucial role in degrading the death cases. Dermatologists diagnose melanoma from dermoscopic images, but it is in a bond with the clinician's expertise level, where the diagnosis accuracy varies from 24\% to 77\% \cite{tran2005assessing}. Therefore, using the automated segmentation system is unquestionable to minimize the diagnostic erroneous that is dependent on the diverse visual interpretation of dermatologists and their experience. However, skin lesion segmentation is challenging due to several dermoscopic image issues, e.g., illumination changes, low contrast of images, distinct texture, position, shapes, color, and boundaries of skin lesions. In addition, visual artifacts in dermoscopic images, such as air bubbles, hair, ruler markers, and blood vessels, make skin cancer segmentation extremely difficult. Medical image segmentation using deep convolutional networks has become a standard de facto. The golden star of the architectures in this field is a U-Net \cite{ronneberger2015u}. The U-Net is a symmetric U-shaped model that can be divided into the decoder and encoder parts as follows: the encoder(or contracting path) gradually applies the successive convolutional layers followed by the downsampling operation to embed the input data into a high dimensional space while reducing the spatial dimension. On the contrary, the decoding path applies the convolution blocks to gradually reduce the feature dimension and reconstruct the spatial dimension by the use of up-sampling operations. Meanwhile, to compensate for the loss of spatial information caused by the downsampling operation, a skip connection path is incorporated to send a copy of low-level features derived from the encoding path. 

The simplicity, outstanding performance, and modular design of U-Net, further motivated researchers to develop several extensions of this network, where the main modification usually addresses the skip connection path to encourage feature re-usability, e.g., U-Net++ \cite{zhou2018unet++}, ferequency attention U-Net \cite{azad2021texture,azad2021deep} U$^{2}$-Net \cite{qin2020u2}, H-DenseUNet \cite{li2018h}, BCDU-Net \cite{azad2019bi}. In H-DenseUNet, Li et al. \cite{li2018h} replaced the original U-Net encoder with the residual network and dense skip connections to extract more complex features. 
Azad et al. \cite{azad2019bi} further improved the skip connection path by incorporating the LSTM module for a non-linear feature representation and demonstrated a significant improvement in the skin lesion segmentation challenge.
Later studies \cite{zhou2018unet++,qin2020u2} focused more on carefully designing skip connection procedures in a multi-scale form to provide highly flexible feature fusion schemes by aggregating features of varying semantic scales with redesigned skip connections. Generally, by considering neural networks in more detail, more semantic representations can be achieved, so  Zhou et al. in U-Net++ \cite{zhou2018unet++} takes the advantage of this idea by using a dense flow of semantic information from the encoder to the decoder. U$^{2}$-Net \cite{qin2020u2} exploits a residual U-Net block design instead of sequential convolutional layers in each hierarchical block of naive U-Net architecture's encoder and decoder paths. This approach benefits from capturing the multi-scale contextual information and directly aggregates it into the inter-block feature mapping of the decoder path to mitigate the loss of fine details features caused by direct up-sampling with large scales. 

The common characteristic of the aforementioned networks is that they are heavily based on the CNN structure, which suffers from a weak global representation, due to the local receptive field \cite{heidari2022hiformer,azad2022transdeeplab}. Hence, CNN-based approaches usually fail to model structural and boundary information preserved in medical images, such as lesion structure \cite{feyjie2020semi}. To tackle the problem of limited receptive fields and to model the global representation, a self-attention mechanism was proposed. Attention U-Net by Oktay et al. \cite{oktay2018attention} was the first approach that investigated the self-attention paradigm in the U-shape structure for medical image segmentation. They proposed an image-grid-based gating module that comprises well-known skip connections to let signals pass through, and to capture the gradient of relevant localization information from the encoder path before it merges with decoder features on the same scale. This strategy makes the model adjust itself to a particular object segmentation task.

TransUNet \cite{chen2021transunet} is one of the early studies that imposes a Transformer on a U-shaped architecture. They proved that Transformer based models present better results than CNN-based self-attention methods. TransUNet utilizes a hybrid CNN-Transformer architecture to leverage detailed high-resolution spatial information from CNN features and the global context encoded by Transformers in the encoder path \cite{chen2021transunet}. An extension of this network is further designed \cite{azad2022transnorm} for the skin lesion segmentation task by incorporating the attention mechanism into the Transformer skip connections. Even though TransUNet provides quite reasonable results, the model suffers from being dependent on CNNs hierarchical feature extraction. To model the network design in a pure transformer manner, Cao et al. \cite{cao2021swin} proposed the Swin U-Net model. This structure uses the Swin Transformer \cite{liu2021swin} blocks to model the U-Net architecture without any convolutional operation.

In this paper, we propose to further enhance the classical concatenation utilized by the operation in the skip connection path of the pure transformer-based approach by imposing attention weights calculated in the encoder path to highlight informative tokens. We also propose a cross-contextual attention method to re-calibrate the extracted feature set. As opposed to \cite{reza2022contextual}, which uses a transformer block to produce the attention, our method uses the already calculated attention map from the encoder part which does not impose any extra memory/computational burden. Furthermore, the Transnorm \cite{azad2022transnorm} approach utilizes a skip connection between the bottleneck and the decoder paths, which can degrade the low-resolution information. In contrast, our approach applies the attention mechanism in each encoder/decoder scale to model the multi-resolution feature representation. Our contributions are as follows:

\noindent$\bullet$~Cross attention mechanism to enhance feature description on the skip connection path. 

\noindent$\bullet$~Imposing attention weight derived from the encoder path to induce spatial attention mechanism. 

\noindent$\bullet$~State-of-the-art results on three public skin lesion segmentation datasets along with publicly available implementation source code via \href{https://github.com/NITR098/AttSwinUNet}{\textcolor{red}{GitHub}}.

\section{Proposed method} \label{sec:method}
Inspired by the recent success of the Swin U-Net model \cite{cao2021swin}, we propose Att-SwinU-Net, an attention-based Swin U-Net extension, for skin lesion segmentation. Our design offers a two-level attention mechanism where the first level integrates the attention weight transferred from the encoder blocks to highlight the important tokens (spatial attention). In contrast, the second attention level focuses more on pair-wise token fusion for feature recalibration (cross-contextual attention). In the following subsections, we will elaborate on our model in more detail. 

\subsection{Swin Transformer block} \label{sec:swin}
Our Swin Transformer block is built based on shifted window multi-head self-attention (MSA). The Swin block is composed of two successive transformer blocks as in \cite{dosovitskiy2020image}, but each multi-head attention is replaced with a window-based attention module; thus, each Swin sub-blocks consists of Layer Norm(LN), residual connection, and 2-layer MLP with GELU non-linearity. Accordingly, the Swin Transformer block is formulated as
\begin{align}
		\hat{\mathbf{z}}^{l}&=\text{W-MSA}\left(\mathbf{L N}\left(\mathbf{z}^{l-1}\right)\right)+\mathbf{z}^{l-1} \nonumber\\
		\mathbf{z}^{l}&=\text{MLP}\left(\text{LN}\left(\hat{\mathbf{z}}^{l}\right)\right)+\hat{\mathbf{z}}^{l} \nonumber\\
		\hat{\mathbf{z}}^{l+1}&=\text{WS-MSA}\left(\mathbf{L N}\left(\mathbf{z}^{l}\right)\right)+\mathbf{z}^{l} \nonumber \\ 
		\mathbf{z}^{l+1}&=\text{MLP}\left(\mathbf{LN}\left(\hat{\mathbf{z}}^{l+1}\right)\right)+\hat{\mathbf{z}}^{l+1} \label{eq:swinequation}
\end{align}
where ${\hat{\mathbf{z}}^{\left( \bullet  \right)}}$  and ${\mathbf{z}^{\left(  \bullet  \right)}}$ denote the output of (S)W-MSA modules and the MLP module of the corresponding block. Finally, attention with respect to relative position $B \in {\mathbb{R}^{{M^2} \times {M^2}}}$ for each head is calculated by
\begin{align}
	\text{Attention}(Q, K, V)=\text{softmax}\left(Q K^{T} / \sqrt{d}+B\right) V \label{eq:mainatteq}
\end{align}
where $Q,K,V \in {\mathbb{R}^{{M^2} \times d}}$ represent query, key and value matrices. $M^2$ and $d$ denote the number of patches in a single window and the projection dimension, respectively. Furthermore, values in bias matrix $B$ are taken from the smaller-sized bias matrix $\hat{B} \in  \mathbb{R} ^{(2M-1)\times(2M-1)}$ as proposed in \cite{liu2021swin}.

\subsection{Encoding Path} \label{sec:encoder}
The encoder module in our design follows the same structure as presented in \cite{cao2021swin} and applies stacked Swin Transformer blocks to embed the input image into a latent space. To this end, three Swin Transformer blocks are utilized to gradually reduce the spatial dimension while increasing the representation dimension. More specifically, after each Swin Transformer block, we use the patch merging layer to merge neighboring patches. The patch merging layer concatenates each $2\times2$ neighbor patch with dimension  $C$  and constructs one patch with dimension $4C$. Following this process, the feature dimension grows by a factor of 4; a linear layer is applied to this patch to decrease the growth factor by 2. Ultimately, the model down-samples the spatial representation while up-sampling the channel representation.

\subsection{Decoding Path} \label{sec:decoder}
Following the symmetrical structure of the U-Net model, the decoder module applies three Swin Transformers blocks to reconstruct the prediction mask. To gradually increase the spatial dimensions while reducing the feature dimensions, we replaced the patch merging layer with the patching expanding layer. The patch expanding layer applies a linear layer on a bottleneck's output ($\frac{W}{32} \times \frac{H}{32} \times8C$) to upsample a channel dimension by factor 2. Then the results representation is rearranged ($\frac{W}{32} \times \frac{H}{32} \times16C$) to finally downsample the channel features by factor 4 and the spatial dimension by factor 2 ($\frac{W}{16} \times \frac{H}{16} \times4C$). This process continues to reconstruct the prediction mask $Y’^{H \times W}$. 

\subsection{Cross Attention Mechanism} \label{sec:cross-attention}
\begin{figure}[h]
	\centering
	\includegraphics[width=\linewidth]{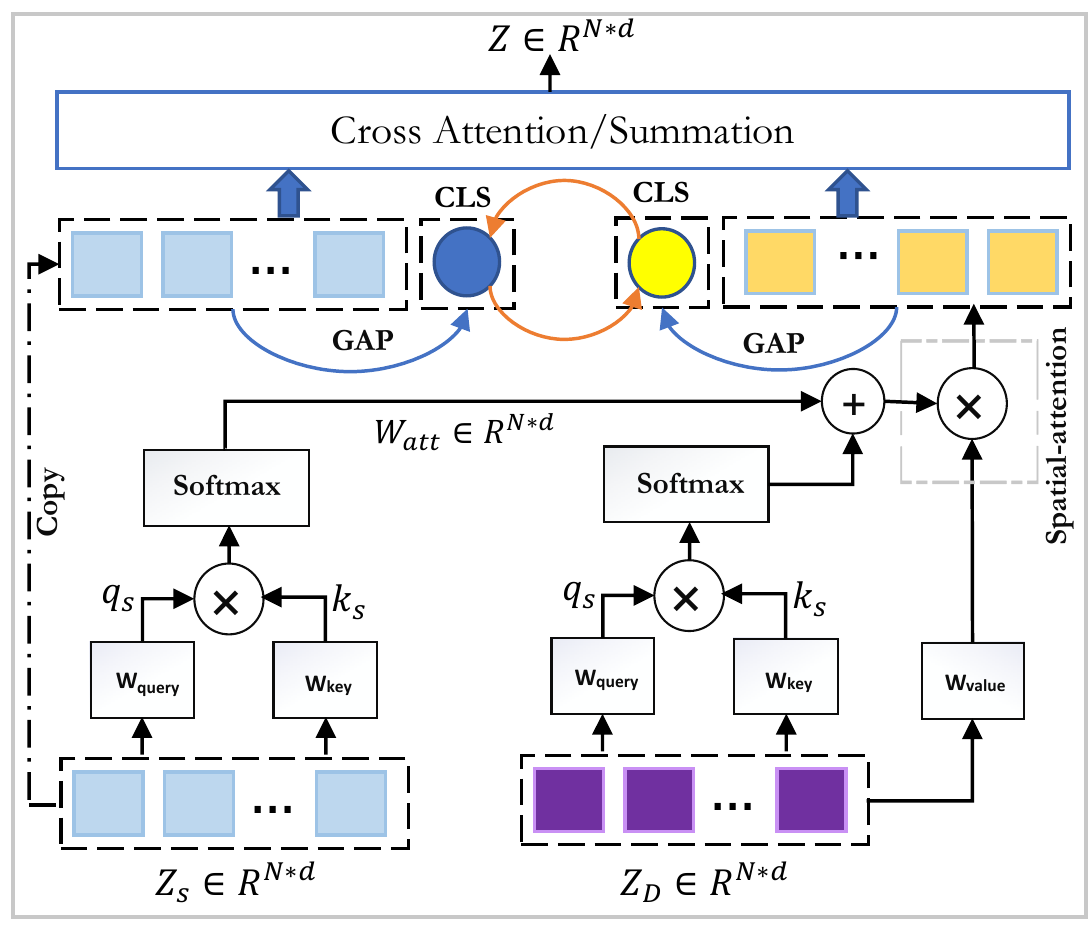}
	\caption{General overview of the proposed attention module to perform spatial and channel attention on the skip connection.}
	\label{fig:proposedmethod}
\end{figure}

The goal of the skip connection path incorporated in the U-Net design is to provide low-level features for the decoding path, which is crucial for localization purposes.  Several extensions of the U-Net model have been proposed in the literature to enhance the skip connection path \cite{azad2019bi,oktay2018attention}. Similarly in this work, we aim to enhance the feature fusion schema in the skip connection section by proposing a two-level attention mechanism. In our design (\autoref{fig:proposedmethod}), first, we apply the spatial normalization mechanism. To this end, the attention weight ($W_{att}$) generated inside the Swin Transformer block of each encoder block passes through the skip connection section to provide a surrogate signal for selectively emphasizing the more informative tokens. Thus, by summing the weight of attention ($W_{att} = \text{softmax}\left(Q_e K_e^{T} / \sqrt{d}+B\right) $) from the encoding path into the decoding path, we guide the network to better model the localization importance (i.e. where to look). According to Eq.\eqref{eq:mainatteq} the attention values in $i$-th scale of the decoder path are as follows:

{\small
\begin{align}
	\text{Att}^{(i)}(Q_d, K_d, V_d)&=(\text{softmax}\left(Q_d K_d^{T} / \sqrt{d}+B\right)+W_{att})V_d \nonumber \\
	s.t. & \quad i \in \{1,2,3\}
\end{align}
}

\noindent where the $\text{Att}^{(i)}(Q_d, K_d, V_d)$, indicates the attention weight calculated in the $i$-th decoder path. The attention mechanism in all Swin Transformer blocks in the decoding path uses this weight calculation technique to perform the weight update in the corresponding scale. In the second step, we model the interaction between two series ($Z_{S}$ and $Z_{D}$) to recalibrate the generated features. To this end, we first average the token representation on each series to produce a global representation, $\mathbf{Z}_{D_{globe}}= \sum Z_{D}/N$, where $N$ indicates the number of tokens. The resulting global representation is then fused with the tokens from the second series 
$\mathbf{Z'}_{D}= [\mathbf{Z}_{S_{globe}}\| Z_{D}]$
and passes through the Swin attention block to re-consider the feature recalibration based on the new global token injected into the sequence (similar to \cite{chen2021crossvit}):

\begin{equation}
\begin{gathered}
\mathbf{q}=\mathbf{Z}_{D_{globe}} \mathbf{W}_{j}, \quad
\mathbf{k}=\mathbf{Z'}_{D} \mathbf{W}_{k}, \quad
\mathbf{v}=\mathbf{Z'}_{D} \mathbf{W}_{l} \\
\mathbf{A}=\text{softmax}\left(\mathbf{q} \mathbf{k}^{T} / \sqrt{C / h}\right), \quad \mathrm{CA}\left(\mathbf{Z'}_{D}\right)=\mathbf{A} \mathbf{v}
\end{gathered}
\end{equation}

\noindent Here,  $\mathbf{W}_{j}$, $\mathbf{W}_{k}$, $\mathbf{W}_{l} \in  {\mathbb{R}^{C \times (C/h)}}$ show the trainable parameters, while $C$ and $h$ indicate the dimension of the embedding space and a number of heads, respectively. The $\text{CA}$ indicates the cross attention that we calculated. This strategy provides a unique way to formulate interaction between the encoder and decoder series in a non-linear fashion.

\section{Experiments} \label{sec:experiments}
Our proposed contributions are based on the modification of the Swin U-Net design; therefore, we use the original implementation of this network (using the PyTorch library) and incorporate our attention mechanism. The Adam optimization with a learning rate of 1e-4 and batch size 24 is utilized to train the network on each dataset for 100 epochs. A single Nvidia RTX 3090 GPU is used to train all models. 
We evaluated our proposed network on three skin lesion segmentation tasks, namely \textit{ISIC 2017} \cite{codella2018skin}, \textit{ISIC 2018} \cite{codella2019skin}, and \textit{PH$^2$} \cite{mendoncca2013ph} datasets. To perform the training and evaluation process, we follow the same pre-processing criteria presented in \cite{azad2022transnorm}. However, instead of an image resolution of $256 \times 256$ we use 
$224 \times 224$ to fit the image into the Transformer model. 

\begin{table*}[t]
\centering
\caption{Performance comparison of the proposed method against the SOTA approaches on skin lesion segmentation task.} \label{tab:quantitative}
\resizebox{\textwidth}{!}{
    \begin{tabular}{l||cccc||cccc||cccc}
    \hline 
    \multirow{2}{*}{\textbf{Methods}} & \multicolumn{4}{c||}{\textit{ISIC 2017}} & \multicolumn{4}{c||}{\textit{ISIC 2018}} & \multicolumn{4}{c}{\textit{PH$^2$}} \\ \cline{2-5} \cline{6-9} \cline{10-13}
    & \textbf{DSC} & \textbf{SE} & \textbf{SP}&\textbf{ACC} & \textbf{DSC} & \textbf{SE} & \textbf{SP}&\textbf{ACC}  & \textbf{DSC} & \textbf{SE} & \textbf{SP}&\textbf{ACC} \\ \hline
    U-Net \cite{ronneberger2015u} & 0.8159 & 0.8172 & 0.9680 & 0.9164 & 0.8545 & 0.8800 & 0.9697 &  0.9404 & 0.8936 & 0.9125 & 0.9588 & 0.9233 \\
    Att U-Net \cite{oktay2018attention} & 0.8082 & 0.7998 & 0.9776 & 0.9145& 0.8566 & 0.8674 & \textbf{0.9863} & 0.9376& 0.9003 & 0.9205 & 0.9640 & 0.9276 \\
    TransUNet \cite{chen2021transunet} &0.8123&0.8263&0.9577&0.9207 & 0.8499 & 0.8578 & 0.9653 & 0.9452 & 0.8840&0.9063&0.9427&0.9200 \\
    MCGU-Net \cite{asadi2020multi}  &  0.8927 & 0.8502 & 0.9855 &  0.9570 & 0.895 & 0.848 & 0.986 & 0.955 & 0.9263 & 0.8322 & 0.9714  & 0.9537 \\
    MedT \cite{valanarasu2021medical} & 0.8037 & 0.8064 & 0.9546 & 0.9090 & 0.8389 & 0.8252 & 0.9637 & 0.9358& 0.9122 & 0.8472 & 0.9657  & 0.9416\\
    FAT-Net \cite{wu2022fat} &0.8500 & 0.8392 & 0.9725 & 0.9326& 0.8903 & \textbf{0.9100} & 0.9699 & 0.9578& 0.9440 & \textbf{0.9441} & 0.9741 & 0.9703 \\
    TMU-Net \cite{reza2022contextual} & 0.9164 & 0.9128 & 0.9789 &  0.9660 & 0.9059 & 0.9038 & 0.9746 & 0.9603 & 0.9414 & 0.9395 & 0.9756 & 0.9647 \\
    Swin\,U-Net \cite{cao2021swin}& 0.9183    & 0.9142    & 0.9798    &  \textbf{0.9701} & 0.8946  & 0.9056  & 0.9798 &  0.9645 & 0.9449  & 0.9410  & 0.9564 &  0.9678 \\
    TransNorm \cite{azad2022transnorm} & 0.8933 & 0.8532 & \textbf{0.9859} & 0.9582 & 0.8951 & 0.8750 & 0.9790 & 0.9580 & 0.9437 & 0.9438 & \textbf{0.9810} & \textbf{0.9723} \\
    \hline
    \textbf{Proposed Method} & \textbf{0.9240} & \textbf{0.9246} & 0.9794  &  0.9656  & \textbf{0.9105} & 0.9089 & 0.9807  &  \textbf{0.9668} & \textbf{0.9504} & 0.9439 & 0.9576 & 0.9685 \\ \hline
    \end{tabular}
}
\end{table*}

\subsection{Quantitative and Qualitative results} \label{sec:results}
\begin{figure}[h]
\centering
\resizebox{\columnwidth}{!}{
\begin{tabular}{@{} *{6}c @{}}
\includegraphics[width=0.16\textwidth]{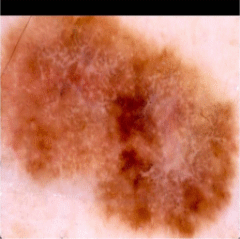} &
\includegraphics[width=0.16\textwidth]{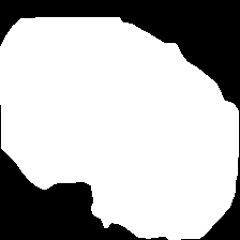} &
\includegraphics[width=0.16\textwidth]{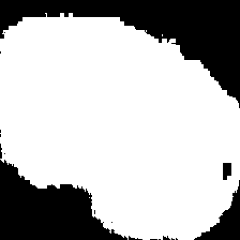} &
\includegraphics[width=0.16\textwidth]{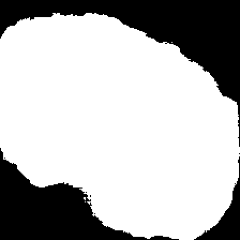} &
\includegraphics[width=0.16\textwidth]{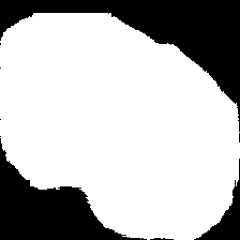} \\
\includegraphics[width=0.16\textwidth]{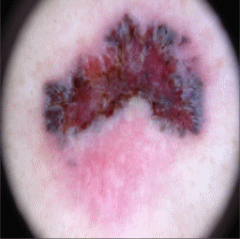} &
\includegraphics[width=0.16\textwidth]{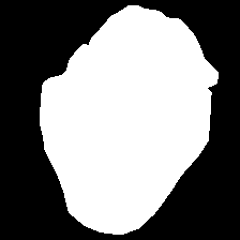} &
\includegraphics[width=0.16\textwidth]{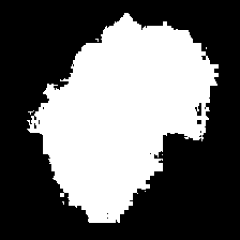} &
\includegraphics[width=0.16\textwidth]{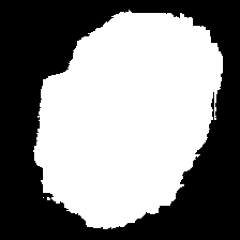} &
\includegraphics[width=0.16\textwidth]{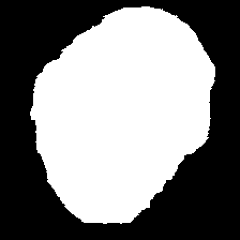} \\
{\small (a)} & {\small (b)} & {\small (c)} & {\small (d)} & {\small(e)}
\end{tabular}
}
\caption{Visual comparisons of different methods on \textit{ISIC2017} dataset. (a) Input image. (b) Ground Truth. (c) TransUNet \cite{chen2021transunet}. 
(d) Swin U-Net \cite{cao2021swin}. (e) Proposed method.} \label{fig:visualcomparison}
\end{figure}

Comparative results of the proposed method on all three skin lesion segmentation datasets are presented in \autoref{tab:quantitative}. We employed different evaluation metrics to provide comprehensive evaluation criteria. According to these results, the proposed method clearly outperforms both CNN and Transformer-based approaches. Overall, our proposed complement modules to Swin U-Net achieve a marked performance across all datasets and evaluation metrics. We also observed that our proposed methods outperform the other Transformer-based peers \cite{wu2022fat,valanarasu2021medical,chen2021transunet} in nearly all skin segmentation benchmarks. Furthermore, we provided a qualitative visualization of the skin lesion segmentation results in \autoref{fig:visualcomparison}. Our proposed method provides smoother segmentation results than others, especially in comparison to TransUNet \cite{chen2021transunet}, which tends to have under-segmentation problems that the locality of convolution operation may cause. In addition, the attention mechanism incorporated in our method slightly reduces the false segmentation results obtained by the Swin U-Net model.

\subsection{Ablation study} \label{sec:ablation}
To explore the influence of different settings on the model performance, we performed ablation analyses on the number of usages of our proposed attention strategy in skip connections, input sizes, model scales, and the presence of the proposed attention mechanism presented. To begin with, we investigated the effect of incorporating our proposed attention at 1/4, 1/8, and 1/16 resolution scales. By changing the number of presence of our proposed module in skip connections from $i \in \left\{ {1, 2 ,3} \right\}$ we explored their effect on our model. \autoref{tab:ablationstudy} demonstrates that the segmentation performance improves when the number of our proposed modules in the skip connections increases. They support the effectiveness of our attention module in capturing a rich representation. We also analyzed the effect of increasing the input size to $384\times384$ in \autoref{tab:ablationstudy}. The results obtained show that with increasing input size the segmentation results improve slightly, but it incurs a high computational cost. In another setting, we explored the impact of model scale growth. It is evident from \autoref{tab:ablationstudy} that the model barely improves its performance due to over-parameterization. Last but not least, we conducted experiments on the proposed attention mechanism to separately analyze the effect of spatial and channel attention steps. Results are presented in \autoref{tab:ablationstudy}, which indicates that each module contributes to the model generalization performance. 

\begin{table}[!thb]
	\centering
	\caption{Ablation study on the \textit{ISIC17} dataset.} \label{tab:ablationstudy}
    \resizebox{\columnwidth}{!}{
	\begin{tabular}{l c c c c} \toprule
		\textbf{Setting} &  \textbf{DSC} & \textbf{SE} & \textbf{SP} & \textbf{ACC} \\ \cmidrule{1-5}
		Using 1 skip connection  & 0.8987 & 0.8767 & 0.9576 & 0.9534 \\
		Using 2 skip connection  & 0.9178 & 0.9147 & 0.9708 & 0.9613 \\
		Using 3 skip connection  & 0.9240 & 0.9246 & 0.9794 & 0.9656 \\
		\cmidrule{1-5}
		Input image size~$384 \times 384$ & 0.9249 & 0.9265 & 0.9799 & 0.9659 \\
		\cmidrule{1-5}
		Large Model   & 0.9245 & 0.9260 & 0.9783 & 0.9660 \\		
		\cmidrule{1-5}
		Eliminating the spatial attention module & 0.9201 & 0.9226 & 0.9616 & 0.9631 \\
		Eliminating the cross contextual attention module & 0.9196 & 0.9104 & 0.9689 & 0.9622 \\
		\bottomrule
	\end{tabular}
    }
\end{table}

\section{Conclusion} \label{sec:conclusion}
This work proposes an attention mechanism to enhance the performance of the recently proposed Swin U-Net model. Our attention module proposes a two-level attention operation, wherein the first step integrates the attention weight transferred from the encoder blocks to highlight the important tokens while the second attention level is designed to perform a cross-contextual attention mechanism. The results delivered in this paper support our claim by improvements over many architectures. \\
\textbf{Compliance with Ethical Standards}: No human or animal subject was used in this study. 

\bibliographystyle{IEEEbib}
\bibliography{refs}

\end{document}